\documentclass{ws-p8-50x6-00}

\newcommand{\lsim}{\mathrel{\rlap{\lower4pt\hbox{\hskip0pt$\sim$}}
\raise1pt\hbox{$<$}}}
\newcommand{\gsim}{\mathrel{\rlap{\lower4pt\hbox{\hskip0pt$\sim$}}
\raise1pt\hbox{$>$}}}
\newcommand{\case}[2]{\mbox{\footnotesize $\displaystyle \frac{#1}{#2}$}}

\begin{document}

\title{Calculating the critical exponents of the chiral phase transition}

\author{Pieter Maris}

\address{Department of Physics, Kent State University,
Kent, OH 44242, USA}

\maketitle

\abstracts{
We calculate the critical exponents of the chiral phase transition at
nonzero temperature using the thermal and chiral susceptibilities.  We
show that within a class of confining Dyson--Schwinger equation (DSE)
models the transition is mean field, and that an accurate determination
of the critical exponents requires extremely small values of the
current-quark mass, several order of magnitude smaller than realistic
up- and down-quark masses.  In general, rainbow truncation models of QCD
exhibit mean field exponents as a result of the gap equation's fermion
substructure.}

\section{Introduction}
It is anticipated that the restoration of chiral symmetry, which
accompanies the formation of a quark-gluon plasma at nonzero temperature
$T$, is a second-order phase transition in QCD with 2 light flavours.
Such transitions are characterised by two critical exponents:
$(\beta,\delta)$, which describe the response of the chiral order
parameters, ${\cal X}$, to changes in $T$ and in the current-quark mass,
$m$.  Denoting the critical temperature by $T_c$, and introducing the
reduced-temperature $t:= T/T_c-1$ and reduced mass $h:=m/T$, then
\begin{eqnarray}
\label{aa} {\cal X} \propto (-t)^\beta\,,\;\;\; && t\to 0^-\,,\;h=0\,,\\
\label{ab} {\cal X} \propto h^{1/\delta}\,,\;\;\;&& h\to 0^+\,,\;t=0\,.
\end{eqnarray}
Calculating the critical exponents is an important goal because of the
notion of {\it universality}, which states that their values depend only
on the symmetries and dimensions, but not on the microscopic details of
the theory.

The success of the nonlinear $\sigma$-model in describing
long-wavelength pion dynamics underlies a conjecture\cite{pisarski} that
chiral symmetry restoration at finite $T$ in 2-flavour QCD is in the
same universality class as the 3-dimensional, $N=4$ Heisenberg magnet
($O(4)$ model), with critical exponents:\cite{ofour} $\beta^H=0.38 \pm
.01$, $\delta^H=4.82 \pm .05$.  However, recently it was argued that the
compositeness of QCD's mesons affects the nature of the phase transition
and the Gross--Neveu model was presented as a counterexample to
universality.\cite{kogut} Subsequent studies\cite{stephanov} indicated
that nontrivial $1/N$ corrections are important and that this model has
the same critical exponents as the Ising model, as was argued on the
notion of universality, but only in a scaling region of width $1/N$.

Calculating the exponents $\beta$ and $\delta$ directly\cite{prl} from
Eqs.~(\ref{aa}) and (\ref{ab}) is often difficult because of numerical
noise near the critical temperature.  Another method is to
consider\cite{arnea} the chiral and thermal susceptibilities:
\begin{equation}
\label{defchih}
\chi_h(t,h) := 
\left.\frac{\partial\, {\cal X}(t,h)}
        {\!\!\!\!\!\!\partial h}\right|_{t}\,,\;\;\;
\chi_t(t,h) :=
\left.\frac{\partial\, {\cal X}(t,h)}
        {\!\!\!\!\!\!\partial t}\right|_{h}\,.
\end{equation}
At each $h$, $\chi_i(t,h)$, $i=h,t$, are smooth functions of $t$ with
maxima $\chi_i^{\rm pc}$ at the pseudocritical points $t_{\rm pc}^i$.
Near the critical point $t = 0 = h$ we have
\begin{eqnarray}
\label{pchpct}
&&        t_{\rm pc}^h \propto\,  h^{1/(\beta \delta)}\,
        \propto\,  t_{\rm pc}^t\,,\\
\label{deltaslope}
&& \chi_h^{\rm pc}\; = \; \chi_h(t_{\rm pc}^h,h) \;\propto\; 
        h^{-z_h}\,,\;\;\;
        z_h := 1 - \case{1}{\delta} \,,\\
\label{betaslope}
&& \chi_t^{\rm pc} \; = \; \chi_t(t_{\rm pc}^t,h) \; \propto \; 
        h^{-z_t}\,,\;\;\;
        z_t:= \case{1}{\beta\delta}\,(1-\beta)\,.
\end{eqnarray}
Therefore, by calculating the chiral and thermal susceptibilities and
locating the pseudocritical points, one can determine $T_c$ and the
critical exponents.\cite{arnea,hmr98}

\section{Quark Dyson--Schwinger Equation}
We have analysed\cite{hmr98} $\chi_h(t,h)$ and $\chi_t(t,h)$ in a class
of confining DSE models that underlies many successful phenomenological
applications\cite{pct} at both zero and finite-$(T,\mu)$.\cite{rs99} The
foundation of our study is the renormalised quark DSE
\begin{eqnarray}
S^{-1}(\vec{p},\omega_k) & :=& 
        i\vec{\gamma}\cdot \vec{p} \,A(p^2,\omega_k)
        + i\gamma_4\,\omega_k \,C(p^2,\omega_k) + B(p^2,\omega_k)   \\ 
&= &Z_2^A \,i\vec{\gamma}\cdot \vec{p} 
        + Z_2^C \, i\gamma_4\,\omega_k 
        + Z_4 \,m_R(\zeta) + \Sigma^\prime(\vec{p},\omega_k)\,.
\label{qDSE} 
\end{eqnarray}
Here $\omega_k= (2 k + 1)\,\pi T$ is the fermion Matsubara frequency and
$m_R(\zeta)$ is the current quark mass at the renormalisation point
$\zeta$. The self-energy is
\begin{eqnarray}
\Sigma^\prime(\vec{p},\omega_k) & =& 
        T\sum_{l=-\infty}^\infty \!\int\! \frac{d^3q}{(2\pi)^3}\,
        \case{4}{3}\,g^2\,D_{\mu\nu}(\vec{p}-\vec{q},\Omega_{k-l})\,
        \gamma_\mu S(\vec{q},\omega_l)\Gamma_\nu \,,
\label{regself}
\end{eqnarray}
with $D_{\mu\nu}(\vec{k},\Omega_j)$ the renormalised dressed-gluon
propagator and $\Gamma_\nu$ the renormalised dressed-quark-gluon vertex.
In renormalising the DSE we require
\begin{equation}
\label{subren}
\left.S^{-1}(\vec{p},\omega_0)\right|_{p^2+\omega_0^2=\zeta^2} = 
        i\vec{\gamma}\cdot \vec{p} + i\gamma_4\,\omega_0 + m_R\;.
\end{equation}
Equations~(\ref{qDSE})-(\ref{subren}) define the exact QCD {\it gap
equation}.

We use the rainbow trunctation for the vertex, $\Gamma_\nu =
\gamma_\nu$, which is the leading term in a $1/N_c$-expansion of the
vertex, and consider three models in which the long-range part of the
interaction is an integrable infrared singularity,\cite{mn83} motivated
by $T=0$ studies of the gluon DSE:\cite{pennington}
\begin{eqnarray}
g^2 D_{\mu\nu}(\vec{k},\Omega_j) &=& 
        P_{\mu\nu}^L(\vec{k},\Omega_j) {\cal D}(\vec{k},\Omega_j;m_g) + 
        P_{\mu\nu}^T(\vec{k},\Omega_j) {\cal D}(\vec{k},\Omega_j;0) \,, 
\nonumber\\
\label{delta}
 {\cal D}(\vec{k},\Omega_j;m_g) &:=& 
        2\pi^2 D\,\case{2\pi}{T}\delta_{0\,j} \,\delta^3(\vec{k}) 
        + {\cal D}_{\rm M}(k^2+\Omega_j^2+m^2_g)\,,
\end{eqnarray}
where $P_{44}^T=P_{4i}^T=0$, $P_{ij}^T=\delta_{ij}-k_i k_j/k^2$,
$P_{\mu\nu}^L=\delta_{\mu\nu}-k_\mu k_\nu/(k^2+\Omega^2)-P_{\mu\nu}^T$,
$m_g$ is a Debye mass, and $D$ is a mass-parameter fitted to $m_\pi$
and $f_\pi$ at $T=0$.  We compare the results for 3 different models,
denoted by ${\cal D}_{\rm M}$, ${\rm M} = {\rm A, B, C}$.

One order parameter for dynamical chiral symmetry breaking is the quark
condensate\cite{mr97} $\langle \bar q q\rangle_\zeta^0$.  There are
other, equivalent order parameters and in calculating the chiral and
thermal susceptibilities we employ
\begin{equation}
{\cal X}:= B(p^2=0,\omega_0), \; \; \; 
{\cal X}_C:= \frac{B(p^2=0,\omega_0)}{C(p^2=0,\omega_0)}.
\end{equation}
They should be equivalent and, as we will see, the onset of that
equivalence is a good way to determine the $h$-domain on which
Eqs.~(\ref{pchpct})-(\ref{betaslope}) are valid.  Further, we have
verified numerically that in the chiral limit ($m=0$) and for $t \sim
0$: $f_\pi \propto \langle\bar q q\rangle \propto {\cal X}(t,0)$; i.e.,
that these quantities are all equivalent, {\it bona fide} order
parameters.  It thus follows from the pseudoscalar mass
formula:\cite{mr97} $f_\pi^2\,m_\pi^2 = 2\,m_R(\zeta)\langle\bar q q
\rangle_\zeta^0$, that $m_\pi$ increases with temperature.\cite{prl}

\section{Results}
The first model we consider is an infrared dominant model with ${\cal
D}_{\rm A}(s) \equiv 0$, and the mass-scale $D=0.56\,$GeV$^2$
fixed\cite{mn83} by fitting $\pi$- and $\rho$-meson masses at $T=0$.  A
current-quark mass of $m=12\,$MeV yields \mbox{$m_\pi=140\,{\rm MeV}$}.
The quark DSE obtained with ${\cal D}_A$ is an algebraic equation.
Chiral symmetry restoration is therefore easy to analyse and either
directly, via Eqs.~(\ref{aa}) and (\ref{ab}), or using the
susceptibilities and Eqs.~(\ref{pchpct})-(\ref{betaslope}), it is
straightforward to establish\cite{arnea} that this model has mean field
critical exponents and to determine the critical temperature in
Table~\ref{taba}.  The exponents are unchanged\cite{hmr98} and $T_c$
reduced by $<\,$2\% upon the inclusion of some higher-order
$1/N_c$-corrections to the dressed-quark-gluon vertex.\cite{brs96}

\subsection{Model B: QED-like tail}
To improve the ultraviolet behaviour, we consider a model\cite{prl} with
\begin{equation}
\label{modelfr}
{\cal D}_{\rm B}(s) = \case{16}{9}\pi^2\,
        \frac{1-{\rm e}^{- s /(4m_t^2)}}{s}\,,
\end{equation}
and $D=(8/9) \, m_t^2 $.  The mass-scale $m_t=0.69\,{\rm
GeV}=1/0.29\,{\rm fm}$ marks the boundary between the perturbative and
nonperturbative domains, and was fixed\cite{fr} by requiring a good
description of $\pi$- and $\rho$-meson properties at $T=0$.

\begin{figure}[t]
\epsfxsize=14pc %
\epsfbox{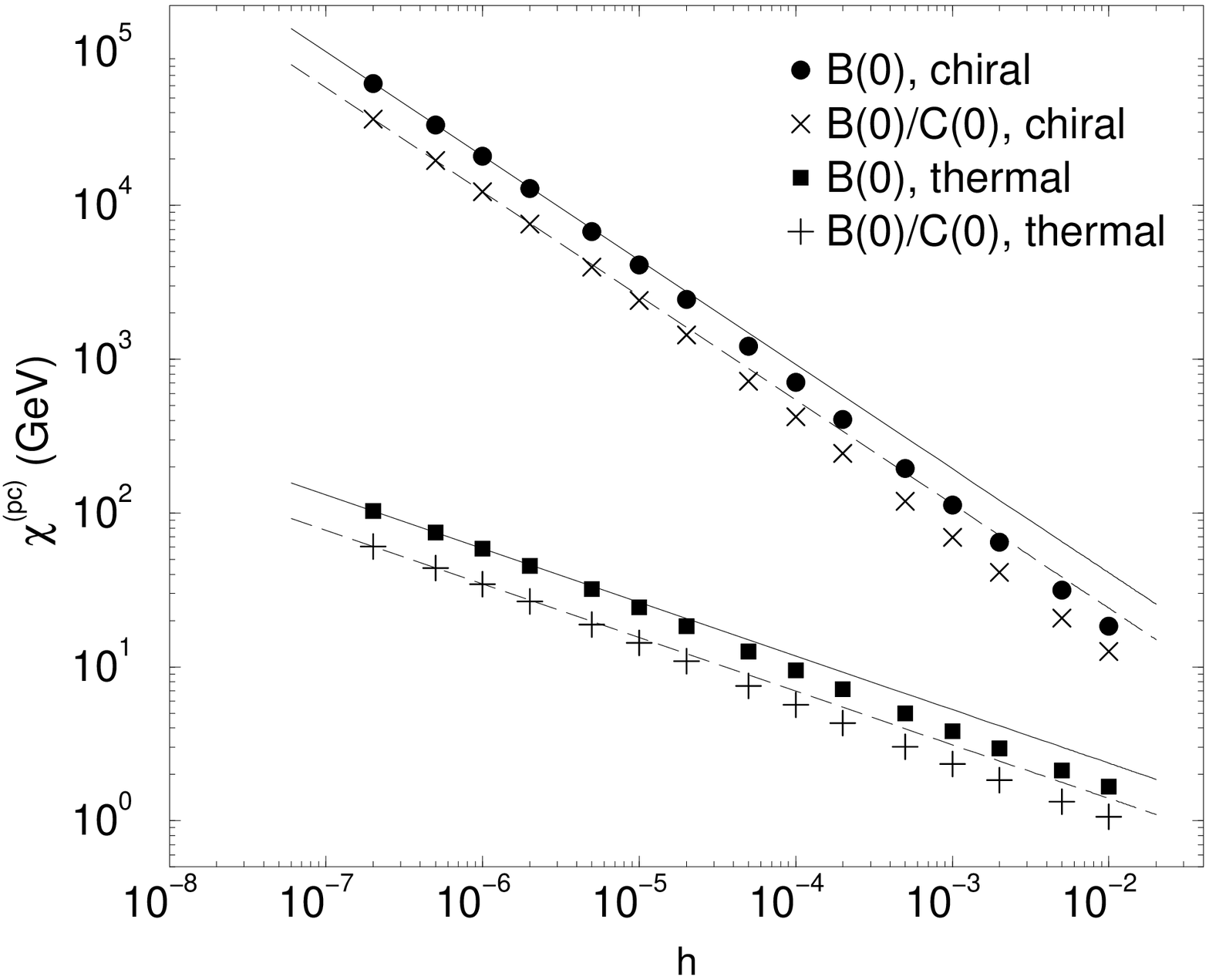}
\epsfxsize=13pc %
\epsfbox{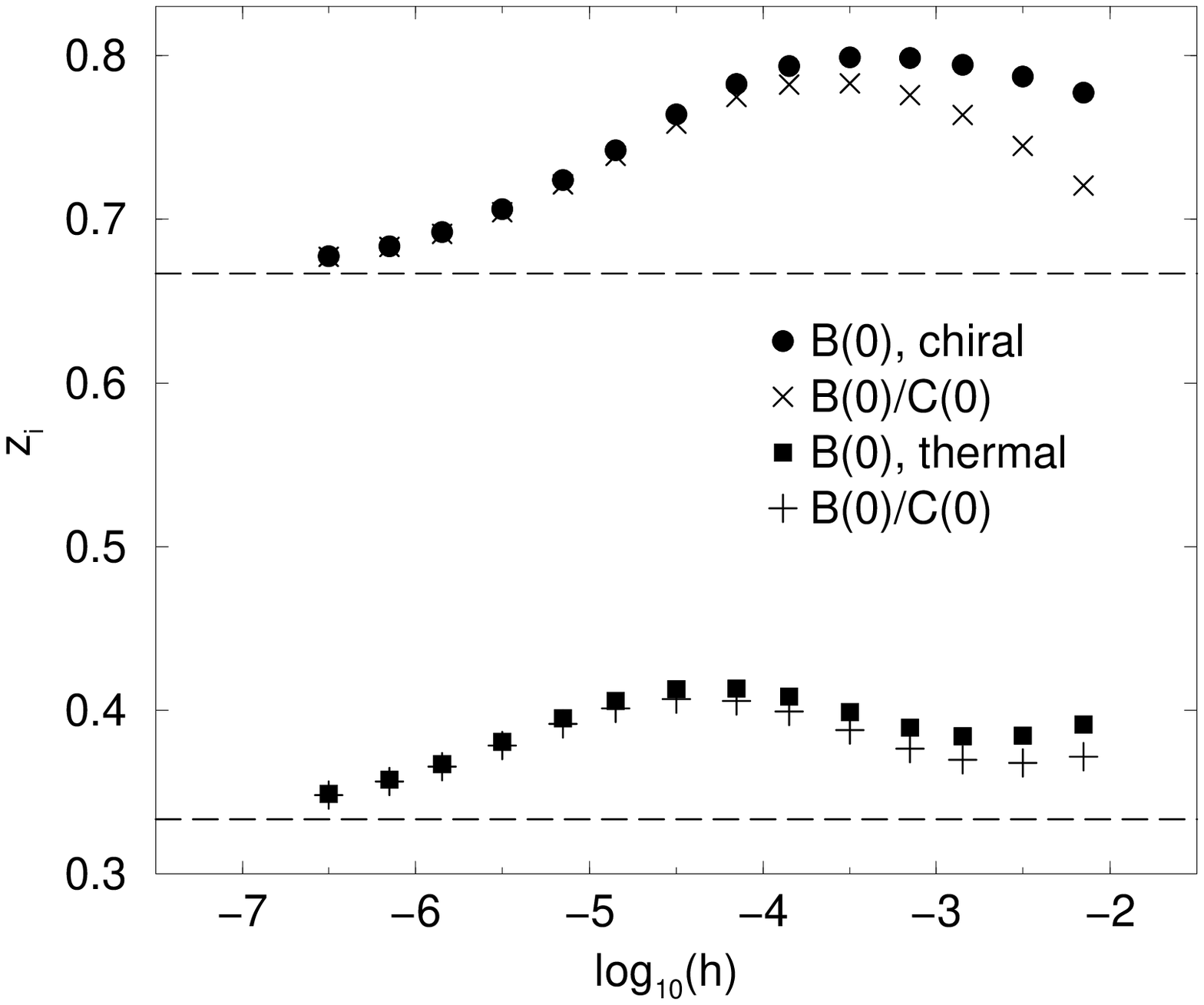}
\caption{(a) $\chi_h^{\rm pc}(h)$ and $\chi_t^{\rm pc}(h)$
calculated in model B.  The slope of the straight lines is given in
Table~\protect\ref{taba} and they are drawn through the smallest
$h$-values; (b) $z_h(h)$ (top) and $z_t(h)$ (bottom) from
Eq.~(\protect\ref{zi}).  The dashed lines are the mean field values:
$z_h=2/3$, $z_t=1/3$.\label{frchi}}
\end{figure}

The quark DSE obtained with this model can be solved numerically and
$\chi_h^{\rm pc}(h)$ and $\chi_t^{\rm pc}(h)$ are depicted in
Fig.~\ref{frchi}(a).
Following Eqs.~(\ref{deltaslope}) and (\ref{betaslope}), the critical
exponents can be determined by defining a local critical exponent as
a function of the reduced mass $h$ for each of the equivalent order
parameters:
\begin{equation}
\label{zi}
z_i(h):= \,-\,h\,\frac{ \partial \ln \chi^{\rm pc}_i}{\partial h}\,,
\end{equation}
see Fig.~\ref{frchi}(b).  $h$ lies in the scaling region when $z_i$ is
independent of the order parameter.  This shows that the scaling
relations are not valid until
\begin{equation}
\label{mfr}
\log_{10} (h/h_u)< -7\,, 
\end{equation}
where $h_u = m_R/T_c$ corresponds to the current-quark mass that gives
\mbox{$m_\pi =140\,{\rm MeV}$} in this model.  The values of $z_h$ and
$z_t$ in Table~\ref{taba} are obtained by a Pad\'e fit to the five
smallest
\begin{table}[b]
\begin{tabular}{c|ccccc}
            &mean field&   A   &  B             & C \\\hline
$T_c$ (MeV) &          &  169  & 174            & 120  \\
$z_h$       &  2/3     & 0.666 & 0.67 $\pm$ 0.01& 0.667 $\pm $ 0.001 \\
$z_t$       &  1/3     & 0.335 & 0.33 $\pm$ 0.02& 0.333 $\pm $ 0.001 
\end{tabular}
\caption{Critical temperature for chiral symmetry restoration and critical
exponents characterising the second-order transition in the three exemplary
models.\label{taba}}
\end{table}
$h$-values in Fig.~\ref{frchi}(b), and extrapolating to $h \rightarrow
0^+$.  The critical temperature is obtained using Eq.~(\ref{pchpct});
its value is insensitive to whether $t_{\rm pc}^h$ or $t_{\rm pc}^t$ is
used and to which of the equivalent order parameters is used.

\subsection{Model C: Logarithmic tail}
Finally, we consider the finite-$T$ extension\cite{hmr98} of a model
which further improves the ultraviolet behaviour, via the inclusion of
the one-loop $\ln$-suppression at $s \gg \Lambda_{\rm QCD}^2$.  Again,
the parameters are fixed at $T=0$ by requiring a good fit to a range of
$\pi$-, $K$-meson properties;\cite{mr97} recent calculations show that
the vector mesons are also described well in this model.\cite{mt99} The
study of chiral symmetry restoration in this model is very similar to
the previous study, with the additional $\ln[s]$-suppression in the
ultraviolet making the numerical analysis easier.  The critical
temperature and exponents are presented in Table~\ref{taba}.  Also in
this case the scaling relations are only valid for very small
current-quark masses: $\log_{10} (h/h_u) < -5$.  These results are
qualitatively, and for the critical exponents quantitatively,
independent of the parameters in this model.  Direct calculation of the
critical exponents using Eqs.~(\ref{aa}) and (\ref{ab}) are in good
agreement with the critical exponents found using the susceptibilities.

\section{Conclusions}
It is clear from Table~\ref{taba} that each of these models is mean
field in nature.  In hindsight that may be not surprising because the
long-range part of the interaction is identical.  However, the models
differ by the manner in which the interactions approach their long-range
limits, and our numerical demonstration of their equivalence required
extremely small values of the current-quark mass, Eq.~(\ref{mfr}).  This
might also be true in QCD; i.e., while $T_c$ is relatively easy to
determine, very small current-quark masses may be necessary to
accurately calculate the critical exponents from the susceptibilities.
In that case, calculation of $\beta$ and $\delta$ via lattice-QCD will
not be easy.  The discrepancies found in recent lattice
calculations\cite{el98} could be a signal of this difficulty.

The class of models we have considered can describe the long-wavelength
dynamics of QCD very well\cite{pct,rs99} in terms of mesons that are
quark-antiquark {\it composites}.  The characteristic feature is the
behaviour of the confining interaction.  It provides a driving term in
the quark DSE proportional to the dressed-quark propagator, which means
that boson Matsubara zero-modes do not influence the critical behaviour
determined from the gap equation.  The class of Coulomb gauge
models\cite{reinhard} also describes mesons as composite particles and
it too exhibits mean field critical exponents.  The long-range part of
the interaction in that class of models corresponds to the regularised
Fourier amplitude of a linearly rising potential. Hence it is not
equivalent to ours in any simple way, except insofar as zero modes do
not influence the gap equation.

The quark DSE is the QCD gap equation and the many equivalent chiral
order parameters are directly related to properties of its solution.  We
have observed that several classes of models exhibit the same (mean
field) critical exponents.  Only in our simplest confining model did we
consider the effect of $1/N_c$-corrections to the quark-gluon vertex,
and in that case the critical exponents were unchanged.  These results
suggest that mean field exponents are a feature of the essential fermion
substructure in the gap equation.  It can likely only be false if
nonperturbative corrections to the vertex are large in the vicinity of
the transition.  In this context the role of mesonic bound states, which
can appear as nonperturbative contributions in the dressed-quark-gluon
vertex, has to be studied in more detail.  This might also give a
nontrivial dependence on the number of fermion flavours, as is
anticipated on universality arguments.

\section*{Acknowledgments}
I would like to thank ECT* and the organizers for providing such a
stimulating environment for this workshop.  This work was done in
collaboration with A.~H\"oll and C.D.~Roberts and funded in part by the
National Science Foundation under grants no.~INT-9603385 and
PHY97-22429, and benefited from the resources of the National Energy
Research Scientific Computing Center.


\begin{thebibliography}{99}
%
\bibitem{pisarski} R. Pisarski and F. Wilczek, Phys. Rev. D{\bf 29}, 338
(1984).
%
\bibitem{ofour}  G. Baker, B. Nickel and D. Meiron, 
Phys. Rev. B{\bf 17}, 1365 (1978).
%
\bibitem{kogut} A. Koci\'c and J. Kogut, Phys. Rev. Lett. {\bf 74}, 3109
(1995).
%
\bibitem{stephanov} J.~Kogut, M.~Stephanov and C.~Strouthos,
Phys.~Rev.~D{\bf 58}, 96001 (1998).
%
\bibitem{prl} A. Bender {\it et al.}, Phys. Rev. Lett. {\bf 77}, 3724 (1996). 
%
\bibitem{arnea} D. Blaschke {\it et al.}, Phys. Rev. C{\bf 58}, 1758 (1998).
%
\bibitem{hmr98} A. H\"oll, P. Maris and C.D. Roberts, 
Phys. Rev. C{\bf 59}, 1751 (1999).
%
\bibitem{pct} P.C. Tandy, Prog. Part. Nucl. Phys. {\bf 39}, 117 (1997).
%
\bibitem{rs99} C.D. Roberts and S. Schmidt, nucl-th/9903075, these proceedings. 
%
\bibitem{mn83} H.J. Munczek and A.M. Nemirovsky, Phys. Rev. D{\bf 28}, 3081
(1983).
%
\bibitem{pennington} N. Brown and M.R. Pennington, Phys. Rev. D{\bf 39},
2723 (1989).
%
\bibitem{mr97} P. Maris and C.D. Roberts, Phys. Rev. C{\bf 56}, 3369
(1997).
%
\bibitem{brs96} A. Bender, C.D. Roberts and L. v. Smekal, Phys. Lett. B 
{\bf 380}, 7 (1996).
%
\bibitem{fr} M.R. Frank and C.D. Roberts, Phys. Rev. C{\bf 53}, 390 (1996).
%
\bibitem{mt99} P. Maris and P.C. Tandy, nucl-th/9905056.
%
\bibitem{el98} E. Laermann, Nucl. Phys. B {\bf 63} (Proc. Suppl.), 114
(1998).
%
\bibitem{reinhard} R. Alkofer, P.A. Amundsen and K. Langfeld, Z. Phys. C
{\bf 42}, 199 (1989).
%
\end{thebibliography}
\end{document}